\documentclass[aps,amssymb,amsmath,prb,eqsecnum,aps,twocolumn,showpacs]{revtex4}
\usepackage[dvips]{graphicx}
\begin{document}
\newcommand{\half}{\frac{1}{2}}
\title{Time-dependent single electron tunneling through a shuttling nano-island}
\author {G. Cohen}
\author{V. Fleurov}
\author{K. Kikoin}
\affiliation{Raymond and Beverly Sackler Faculty of Exact
Sciences, School of Physics and Astronomy, Tel-Aviv University,
Tel-Aviv 69978 Israel}

\begin{abstract}
We offer a general approach to calculation of single-electron
tunneling spectra and conductance of a shuttle oscillating between
two half-metallic leads with fully spin polarized carriers. In
this case the spin-flip processes are completely suppressed and
the problem may be solved by means of canonical transformation,
where the adiabatic component of the tunnel transparency is found
exactly, whereas the non-adiabatic corrections can be taken into
account perturbatively. Time-dependent corrections to the tunnel
conductance of moving shuttle  become noticeable at finite bias in
the vicinity of the even/odd occupation boundary at the Coulomb
diamond diagram.

\end{abstract}
\maketitle

\section{Introduction}

Single electron tunneling (SET) is a salient feature of quantum
transport in nanostructures. The SET phenomenon is observed in
various systems, e.g. quantum dots in a tunnel contact with
metallic electrodes, \cite{Pugla05,Cowre} molecular bridges
between the edges of broken metallic wires, \cite{Park,Yu,Roch}
atoms and molecules absorbed on metallic surfaces in a contact
with the tip of tunnel microscope, \cite{Bode,Otte} etc. The study
of electron tunneling through the nano-object with time-dependent
characteristics is one of the most challenging problems in this
field.

There are several sources of time dependence, which may be realized
in practical devices. The simplest one is the time-dependent gate
voltage $v_g(t)$ applied to the dot. It is well known
\cite{GA,KNG,KKAR} that this time dependence may be converted into
the time dependence of tunnel matrix element. Another possibility is
the nanoelectromechanical shuttling (NEMS),\cite{shut} where the
nano-size island suspended on a pivot\cite{pivot} or attached to a
string\cite{Koenig} oscillates between the leads under the action of
an electro-mechanical force. In case of molecular bridges, the
vibration eigen modes may be the source of the periodical
oscillations of tunneling parameters.\cite{Has,Roch}

Usually the tunneling between metallic leads and such a nanoobject
is accompanied by many-particle Kondo screening effect
\cite{GR,Ng} resulting in specific type of zero-bias anomaly (ZBA)
in tunnel conductance. Modification of Kondo regime because of
periodically modulated in time tunneling rate due to the
center-of-mass oscillations was studied recently in several
papers. If the oscillations are the eigen modes of a nanoobject
(molecule), then the Kondo-peak (zero bias anomaly in tunnel
conductance) may transform into dip due to the destructive
interference with vibrational mode. \cite{Has} In case of
adiabatic motion induced by electromechanical forces (NEMS)
\cite{shut} the Kondo temperature follows the periodical evolution
of the dot position and increases eventually due to effective
reduction of the average distance between the dot and the leads,
\cite{KKSV} which is determined by the mean square displacement of
the dot (in analogy with Debye-Waller effect in scattering
intensity). Non-adiabatic enhancement of  Kondo tunneling through
such moving nanoobject at finite source-drain bias has been
studied recently.\cite{Gok}

In the present paper we consider adiabatic and non-adiabatic
time-dependent effects in conventional cotunneling between
metallic leads due to periodic modulation of lead-dot tunneling
rate. To suppress many-particle Kondo screening effects, one
should consider leads with magnetically polarized electrons. The
tunneling between  the ferromagnetically ordered leads was
discussed recently\cite{Mart1,Bulka,Lopa02,Choi,Tan04,Chin08,Pas}
in the context of Kondo effect. We are interested in the
situation, where the spin-flip cotunneling is completely
suppressed at small lead-dot bias and low temperatures. Such
tunneling regime may be realized in half-metallic ferromagnets,
where the Fermi surface is formed only by the majority spin
electrons, whereas the spectrum of minority spin carries is
gapped. The electronic and magnetic properties of such metallic
compounds are reviewed in Ref. \onlinecite{Kats}. From the point
of view of existing devices, where the leads are formed by
two-dimensional electrons in degenerate semiconductors, the
relevant material for our studies is dilute magnetic semiconductor
(Ga,Mn)As.\cite{Jung} The indirect magnetic exchange between Mn
impurity ions is responsible for the long-range ferromagnetic
order in this material. This exchange is mediated by
spin-polarized carriers near the top of the valence band.  The
tunnel current in the half-metallic regime arises due to the
minority spin hole cotunneling.

We will show that in the absence of Kondo effect the problem of
tunneling through moving nano-island (quantum dot) may be solved
by means of \textit{time-dependent canonical transformation},
which exactly takes into account both adiabatic and non-adiabatic
lead-dot tunneling processes diagonal in the lead indices. The
non-diagonal source-drain cotunneling may be treated by the
canonical transformation method only perturbatively, but the
adiabatic and non-adiabatic contributions into tunnel conductance
may be sorted out also in this case. It will be shown that the
time-dependent contribution into current-voltage characteristics
of moving nano-object is significant near the boundaries of
Coulomb diamonds.

\section{Model}

\begin{figure}[h]
\includegraphics[width=5.5cm,angle=0]{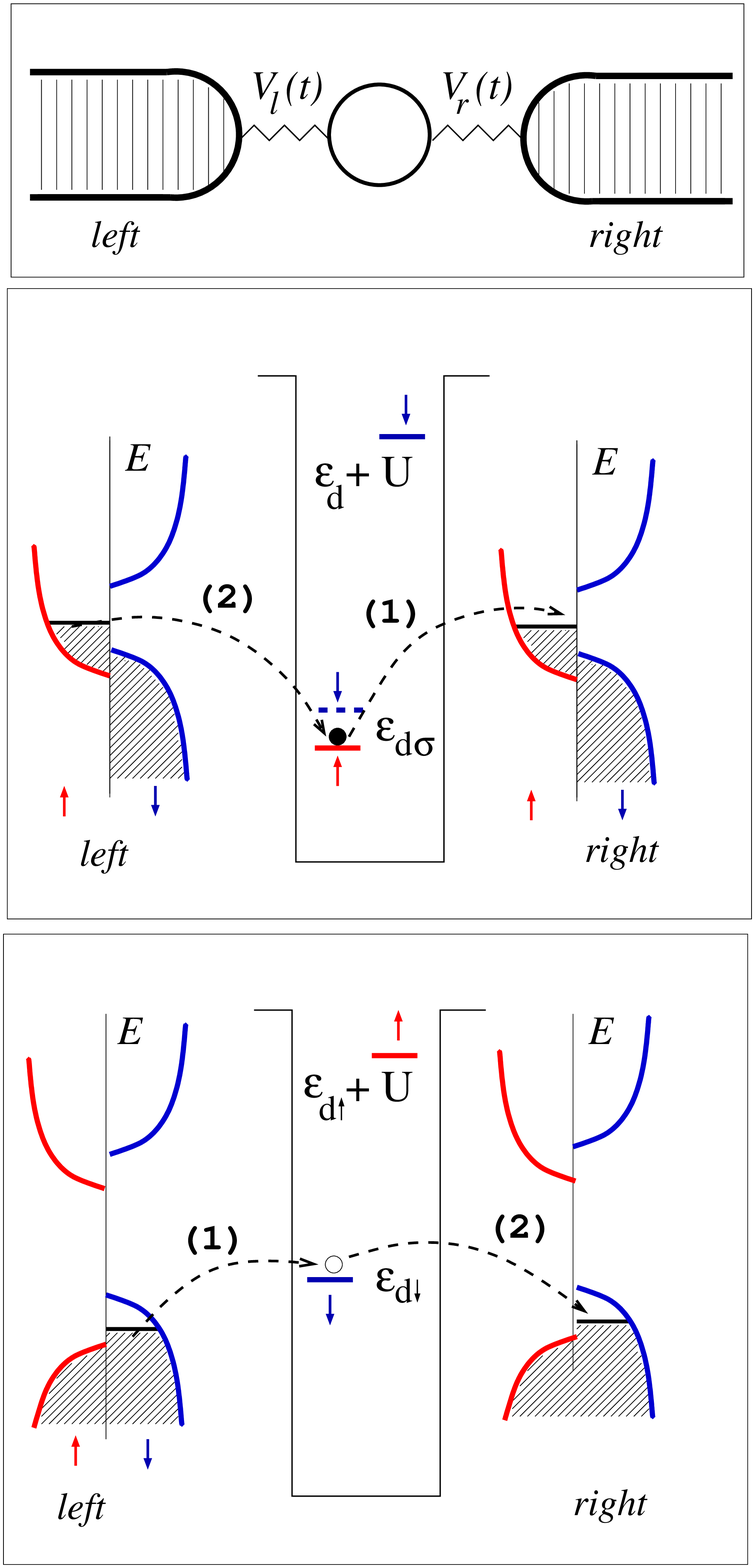}
\caption{Upper panel: model of quantum dot in time-dependent
contact with leads. Middle and lower panels: density of electron
states in the leads, occupied and empty states in the dot for
$N=1$;  mechanisms of spin-polarized electron cotunneling (middle
panel) and hole cotunneling (lower panel) are indicated by arrows.
Figures in parentheses point out the sequence of electron
tunneling acts in cotunneling processes.}\label{f.spol}
\end{figure}

We choose for the realization of ac-driven tunnel conductivity the
simplest model of a nanoobject, which is widely used in the
studies of single-electron tunneling (SET). A nanoobject is
represented in this model by the quantum well with resonance level
$\varepsilon_d$ (see Fig. \ref{f.spol}, upper panel). The SET
regime arises due to the Coulomb blockade effect: addition energy
for the second electron in a singly occupied dot is $\varepsilon_d
+ U \gg \Gamma_j$, where $\Gamma_j$ is the tunneling rate to the
left ($j=l$) and right ($j=r$)lead, and $U$ is the capacitive
energy of the dot. To separate the time-dependent SET from the
Kondo ZBA, we consider tunneling between spin-polarized leads,
where spin-flip processes are inelastic, because the continuum of
electron-hole pairs with the opposite spins in the leads
responsible for the Kondo effect is gapped. Two types of spin
polarized (magnetically ordered) metallic leads are presented
schematically in Fig. \ref{f.spol}. In the middle panel the leads
are formed by "half-metallic ferromagnets"\cite{Kats} with gapped
spectrum of minority spin carries.  In the lower panel
characteristic for $p$-type degenerate dilute magnetic
semiconductors\cite{Jung} the carriers are the minority spin
holes.

Virtual tunneling results in a shift of level positions in QD.
This renormalization ("Friedel shift") is also spin dependent. As
a result the spin polarization of QD adjusts to that of the
ferromagnetic lead (see calculations below). All Kondo processes
are quenched in this regime.

The Anderson Hamiltonian modeling SET has the form
\begin{equation}\label{And01}
H=H_{\rm d} + H_{\rm b} + H_{\rm tun},
\end{equation}
where the terms
\begin{equation}
 H_{\rm d}=\sum\limits_{\Lambda}
E_{\Lambda}|\Lambda\rangle \langle \Lambda|,\ \ H_{\rm b}
=\sum_{j=l,r}\sum\limits_{k\sigma} \epsilon^{}_{jk\sigma}
a_{jk\sigma}^\dagger a^{}_{jk\sigma}
\end{equation}
describe the electron states in the isolated dot and two metallic
(semiconductor) leads, respectively. We write $H_{\rm d}$ in terms
of its eigenstates $|\Lambda\rangle$ (Hubbard representation). This
trick allows one to take all intradot interactions into account
exactly even when the contact with the leads is switched on. The
tunneling  Hamiltonian
$$
H_{\rm tun} = \sum\limits_{jk\sigma}(V_{jk}d^\dagger_{\sigma}
a^{}_{jk\sigma} + \mathrm{H.c.}).
$$
may be rewritten in the Hubbard representation by expanding the
creation operator $d^\dag_\sigma$ in terms of the configuration
change operators $X^{\Lambda \lambda} = |\Lambda\rangle \langle
\lambda|$ which connects the states in adjacent charge sectors
with $N$ and $N-1$ electrons in the dot. We confine ourselves with
the simplest case, where only three charge sectors $N=0,1,2$ are
involved in SET Hamiltonian. Then the Hamiltonian (\ref{And01})
acquires the following form
\begin{eqnarray}\label{ham1}
H &=& \varepsilon_d \sum_\sigma X^{\sigma\sigma} + E_{2}X^{22}
 + \sum_{jk\sigma} \epsilon_{jk\sigma}
a^\dagger_{jk\sigma} a^{}_{jk\sigma} \nonumber \\
& + &\sum_{jk\sigma} [ V_{jk} a^\dagger_{jk\sigma}
(X^{0\sigma}+\sigma X^{\bar\sigma 2}) +{\rm H.c.}].
\end{eqnarray}
Here the quantum numbers $\Lambda,\lambda = 0,\sigma,2$ correspond
to empty, singly and doubly occupied states of QD, respectively,
$E_{2}=2\varepsilon_d+U$ is the energy of doubly occupied QD. The
last term in this Hamiltonian is time-dependent.

\section{Canonical transformation of Anderson Hamiltonian}

Our program is to exclude the tunneling term from the Hamiltonian
(\ref{ham1}) by means of the canonical transformation
\begin{equation}\label{Canon01}
\widetilde{H}=e^S H e^{-S},
\end{equation}
then derive the  tunnel current operator in the new basis and
calculate the tunnel conductance. It was shown in Ref.
\onlinecite{KF79} that this transformation may be performed exactly
in the absence of Coulomb correlation, provided the energy level
$\varepsilon_d$ falls into the energy gap and remains there after
renormalization (Friedel shift). It will be shown below that the
matrix $S$ still may be found in the presence of Coulomb blockade
under the same condition of discreteness of renormalized $d$-level
provided the spin-flip processes are quenched. As may be perceived
from Fig. \ref{f.spol}, this condition is realized for the majority
spin electrons in the case of half-metallic ferromagnet and for the
minority spin holes in the case of $p$-type dilute magnetic
semiconductor.

As usual, the canonical transformation is made by means of the
Baker-Hausdorff expansion
\begin{equation}\label{haus}
e^S H e^{-S} = \sum_{n=0}^\infty \frac{1}{n!}[S,[S,\dots[S,H]\dots]]
\end{equation}
In the non-interacting case, the second quantization operators in
$S$ and $H$ possess Fermi-like commutation relations, the
Hamiltonian $H$ is a quadratic form and the tunnel operator
conserves spin, so the series in the r.h.s. of (\ref{haus}) may be
summed exactly. \cite{KF79} The Coulomb blockade separates the
Hilbert space for the dot electron operators into charge sectors
divided by the energy gaps. As a result these operators lose the
simple Fermi statistics.

We are interested in the strong Coulomb blockade case and start
with the simplest case, where the ground state of the dot
corresponds to $N=1$ in the limit of $U\gg |\varepsilon_d -
\epsilon_F|$, so that the doubly occupied states are completely
suppressed. Then only the states $\Lambda = \sigma,0$ are retained
in the Hamiltonian (\ref{ham1}). In particular, the
anticommutation relation for Hubbard operators mixing the adjacent
sectors $N=0,1$ has the form
\begin{equation}\label{com1}
[X^{\sigma 0}, X^{0\sigma'}]_+ = X^{\sigma\sigma'}+
X^{00}\delta_{\sigma\sigma'},
\end{equation}
which follows from the obvious multiplication rule $X^{\Lambda_1
\lambda_1}X^{\lambda_2 \Lambda_2} = \delta_{\lambda_1\lambda_2}
X^{\Lambda_1\Lambda_2}$. Even disregarding the spin-flip processes
each commutation operation in the expansion (\ref{haus}) generates
operators $K^\sigma=X^{\sigma\sigma}+ X^{00}$. In spite of this, the
Baker-Hausdorff series still can be summed exactly, because these
operators are idempotent, $K^\sigma K^\sigma = K^\sigma$, like
conventional fermion occupation operators $d^\dag_\sigma d_\sigma$.
This summation will be performed in the following subsection.

Time-dependent problem is more complicated because in that case
the canonical transformation should be applied to the operator
\begin{equation}\label{elop}
{\cal L}= -i\hbar\frac{\partial}{\partial t}+ H~.
\end{equation}
In subsection B we will show that the canonical transformation is
operational in this case as well, at least in some important special
cases.

\subsection{Time-independent transformation}
\label{TIT}

In this section we generalize the canonical transformation
proposed for an Anderson model applied to transition metal
impurities in semiconductors.\cite{KF79,KF94} In those
calculations the intra-impurity Coulomb interaction was taken into
account in Hartree approximation. Here we take the Coulomb
blockade term exactly, when summing the series (\ref{haus}), at
least in the charge sectors $N=0,1$. The antihermitian operator
$S$ is looked for in the form
$$
S=\sum_{jk\sigma}(u_{k\alpha} a^\dagger_{jk\sigma} X^{0\sigma} -
u^*_{jk} X^{\sigma 0} a_{jk\sigma})= \sum_\sigma S_\sigma.
$$
If the spin flip processes are neglected, the canonical
transformation is made for each spin projection separately. One
may apply the transformation to the majority spin in the case of
electron tunneling and to the minority spin in the case of hole
tunneling in two models introduced above.

One easily derives the commutation relations
$$[S_\sigma, X^{\sigma 0} ] =
\sum_{jk}u_{jk\sigma}a^\dagger_{jk\sigma} K^{\sigma} \equiv
C_\sigma^\dagger K^{\sigma},
$$$$
[S_\sigma,a^\dagger_{jk\sigma}] = - u^*_{jk\sigma} X^{\sigma 0}.
$$
To shorten notations, we introduce the quantities $C_\sigma^\dagger
= \sum\limits_{jk\alpha} u_{jk\sigma}a_{jk\sigma}^\dagger$ and
$\gamma_\sigma^2 = \sum\limits_{jk} u_{jk\sigma}u^*_{jk\sigma}$.
Besides, we omit the lead index $j$ and specify the band state by a
single index $k$ characterizing both the lead and the electron wave
number. Using these definitions and the above mentioned idempotency
of operator $K_\sigma$, we obtain the following expressions for the
transformed operators
\begin{equation}\label{oneimpa}
\widetilde X^{\sigma 0}=
$$$$
e^{S_\sigma} X^{\sigma 0} e^{-S_\sigma} = X^{\sigma 0}
\cos\gamma_\sigma + C^\dagger_\sigma K^{\sigma} \gamma_\sigma^{-1}
\sin\gamma_\sigma,
\end{equation}
\begin{equation}\label{oneimpb}
\widetilde a^\dag_{k\sigma} = e^{S_\sigma} a^\dagger_{k\sigma}
e^{-S_\sigma} =
$$$$
a^\dagger_{k\sigma} + u^*_{k\sigma} C^\dagger_{\sigma} K^{\sigma}
\gamma_\sigma^{-2} (\cos\gamma_\sigma -1) - u^*_{k\alpha\sigma}
X^{\sigma 0} \gamma_\sigma^{-1} \sin\gamma_\sigma,
\end{equation}
and
\begin{equation}\label{oneimpc}
e^S C^\dagger_{\sigma}e^{-S} = C^\dagger_{\sigma} +
C^\dagger_{\sigma}K^\sigma (\cos\gamma_\sigma - 1)- X^{\sigma 0}
\gamma_\sigma \sin\gamma_\sigma,
\end{equation}

The tunneling term in the transformed Hamiltonian is eliminated,
provided
\begin{equation}\label{coeff}
u_{k\sigma} = \frac{\gamma_\sigma V_{k}}{(\varepsilon_{k\sigma} -
E_{d\sigma})\tan \gamma_\sigma} \equiv g_{k\sigma} V_k
\end{equation}
with
\begin{equation}\label{deriv}
\tan^2\gamma_\sigma = -\left. \frac{dL_\sigma( \varepsilon)}{d
\varepsilon} \right|_{ \varepsilon = E_{d\sigma}} ~,
\end{equation}
\begin{equation}\label{massop}
L_\sigma(\varepsilon) = \sum_{k} \frac{|V_{k}|^2}{\varepsilon -
\varepsilon_{k\sigma}}.
\end{equation}

Then the transformed Hamiltonian takes the form
\begin{equation}\label{htild}
\tilde H_\sigma =  E_{d\sigma} X^{\sigma\sigma} + \sum_{kk'}
\tilde \varepsilon^\sigma_{kk'} a^\dag_{k\sigma}a^{}_{k'\sigma}
\end{equation}
where the renormalized level position is given by the equation
\begin{equation}\label{etild}
E_{d\sigma} = \frac{\varepsilon_d + Z_\sigma \tan^2\gamma_{\sigma}
- (T_\sigma + T_\sigma^*) \tan\gamma_\sigma}{1 +
\tan^2\gamma_\sigma},
\end{equation}
with $Z_\sigma = \gamma_\sigma^{-2} \sum_{k} \varepsilon_k
u^*_{k\sigma} u_{k\sigma},$ and
$ T_\sigma = \gamma_\sigma^{-1} \sum_k (V_k u^*_{k\sigma})$.
Substitution of Eqs. (\ref{coeff}) -- (\ref{massop}) in
(\ref{etild}) transforms it into the conventional
form\cite{KF79,KF94}
\begin{equation}\label{level}
E_{d\sigma} = \varepsilon_d + L_\sigma(E_{d\sigma}).
\end{equation}
where the self energy $L_\sigma(E)$ has only real part (Friedel
shift), provided the level $E_{d\sigma}$ remains within the gap,
which is the case for $\sigma=\uparrow$ in a configuration of Fig.
\ref{f.spol}, middle panel.

A canonical transformation for the second spin component
$\sigma=\downarrow$ should be done more carefully, because the
bare level $\varepsilon_d$ falls into continuum of spin-down
states. One should be accurate with turning to the thermodynamic
limit, where the sum in the right-hand side of (\ref{massop})
transforms into the integral and acquires the imaginary part thus
making the Hamiltonian non-Hermitian. The recipe is to keep the
spectrum of electrons in the leads discreet when doing the
canonical transformation. Then equation (\ref{level}) has ${\cal
N}+1$ solutions $\varepsilon_i$, where ${\cal N}$ is the number of
state in the valence band $\epsilon^{}_{k\downarrow}$. Using Eqs.
(\ref{massop}) and (\ref{deriv}), the corresponding coefficients
$\gamma_i$ may be found. In accordance with (\ref{oneimpa}), the
factor $\cos\gamma_i$ determines the weight of the $d$-component
of hybridized wave function of dot electron in the state $i$. One
may identify the state $i = m$ producing the maximum value of
$\cos\gamma_i$ with the center of future 'Friedel resonance',
which arises in the thermodynamic limit ${\cal N}\to \infty$. In
this sense the state $E_{d\downarrow}\equiv
\varepsilon_{m\downarrow}$ is formally defined from the equation
$\varepsilon_{m\downarrow} = \varepsilon_d +
L_\downarrow(\varepsilon_{m\downarrow})$. This level is shown by
the dashed line in the middle panel of Fig. \ref{f.spol}. It
corresponds to the bunch of excited states of QD coupled to the
leads with $N=1$ and spin oriented antiparallel to that of the
magnetized leads.

The spectrum of continuous part of the Hamiltonian (\ref{htild})
is determined by the expression
\begin{equation}\label{tildec}
\tilde \varepsilon^\sigma_{kk'} = \varepsilon_{k\sigma}
\delta_{kk'}  + W^\sigma_{kk'}
\end{equation}
containing the "scattering" matrix element which eventually
predetermines the tunnel current.

The general form of this matrix element is
\begin{widetext}
\begin{equation}\label{scatamp}
W^\sigma_{kk'} =
%$$$$
\frac{u_{k\sigma} u^*_{k'\sigma}}{\gamma_\sigma^2} K^\sigma \left[
(2\epsilon_d - \varepsilon_{k\sigma} - \varepsilon_{k'\sigma} )(1 -
\cos\gamma_\sigma)- \frac{T_\sigma + T_\sigma^*}{2}
\frac{2\cos^3\gamma_\sigma - 3 \cos^2\gamma_\sigma +
1}{\cos\gamma_\sigma\sin\gamma_\sigma}\right] +
$$$$
K^\sigma(V_ku^*_{k'\sigma} + u_{k\sigma}V^*_{k'})
\frac{\sin\gamma_\sigma}{\gamma_\sigma}.
\end{equation}
\end{widetext}

Using Eqs. (\ref{coeff}) and (\ref{etild}) we get
$$
T_\sigma = T_\sigma^* = - \frac{1}{\tan \gamma_\sigma}
L_\sigma(E_{d\sigma})
$$
and after some algebraic manipulations the scattering matrix element
is eventually transformed into a quite compact expression
\begin{equation}\label{scatamp-tit}
{\overline{W}}^\sigma_{kk'} =
$$$$
V_{k}V_{k'}^*K^\sigma\left[\left(
\frac{1}{\Delta_{k\sigma}}+\frac{1}{\Delta_{k'\sigma}}\right)R(\gamma_\sigma)
+ \frac{L_\sigma(E_{d\sigma})}{\Delta_{k\sigma}
\Delta_{k'\sigma}}R^2(\gamma_\sigma) \right]
\end{equation}
where
\begin{equation}\label{factor}
R(\gamma_\sigma)= \displaystyle\frac{\sqrt{1 + \tan^2\gamma_\sigma}
- 1}{\tan^2\gamma_\sigma},
\end{equation}
$\Delta_{k\sigma} = E_{d\sigma}-\varepsilon_{k\sigma}$. Equation
(\ref{scatamp-tit}) holds in the static case or, as we will see
below, for adiabatically slow time variations of the tunneling
amplitudes. If we want to study nonadiabatic corrections a more
general formula Eq. (\ref{scatamp}) should be used.

Equation (\ref{scatamp-tit}) generalizes the familiar 2-nd order
expression for the single electron tunneling amplitude through the
QD, which takes into account both renormalization of the energy
level of dot electron (\ref{level}) and reconstruction of the band
continuum (\ref{oneimpb}). Far from the resonance tunneling regime
(in the center of the Coulomb diamond diagram, see e.g. the middle
panel of Fig. \ref{f.spol}) one may neglect the second term in the
square bracket of Eq. (\ref{scatamp-tit}), and the tunneling
matrix element acquires the simple form $
{W}^\sigma_{lk,rk'}=J_{lk,rk'}K^\sigma R(\gamma_\sigma) $, where
the first factor is the off-diagonal matrix element of indirect
exchange between the leads and the dot due to electron
cotunneling, the second and the third factors regulate the
occupation of the dot level and the normalization of electron wave
functions, respectively. The transparency of QD is, of course,
exponentially weak

\begin{figure}[h]
\includegraphics[width=5.5cm,angle=0]{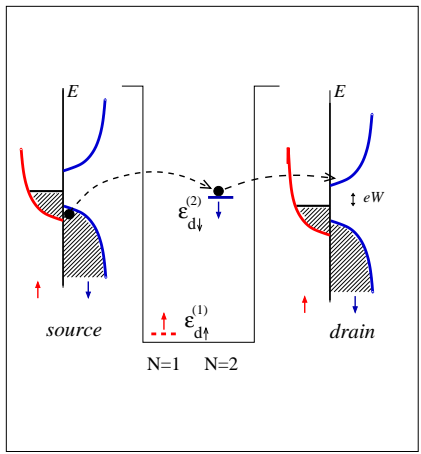}
\caption{Cotunneling mechanism in the resonance regime for second
electron in QD on the boundary of Coulomb window. }\label{f.spul}
\end{figure}

Now we turn to the resonance regime illustrated by Fig.
\ref{f.spul}, where the level $\varepsilon_d^{(2)} = \varepsilon_d +
U$ driven by the gate voltage $v_g$ approaches the level
$\epsilon_F$ from above.
The level $\varepsilon_d^{(2)}$ may be occupied only by a
spin-down electron. Since it falls into the gap of spin-down
density of states in the leads, the canonical transformation
introduced above may be performed in a similar way, provided the
addition energy for the first electron $\varepsilon^{(1)}\approx
\varepsilon_d$ falls deep enough below $\epsilon_F$ and the
corresponding processes are suppressed. In this regime
$\Lambda,\lambda = \sigma,2$, and one should use the commutation
relations
\begin{equation}\label{com2}
[X^{\sigma 2}, X^{2\sigma'}]_+ = X^{\sigma\sigma'}+
X^{22}\delta_{\sigma\sigma'},
\end{equation}
instead of (\ref{com1}). Correspondingly, one should insert
$K^\sigma=X^{\sigma\sigma}+X^{22}$ in equations
\begin{equation}\label{oneimpd}
\widetilde X^{2\bar\sigma }= \sigma X^{2\bar\sigma}
\cos\gamma_\sigma + C^\dagger_\sigma K^{\sigma} \gamma_\sigma^{-1}
\sin\gamma_\sigma,
\end{equation}
\begin{equation}\label{oneimpe}
\widetilde a^\dag_\sigma =
$$$$
a^\dagger_{k\sigma} + u^*_{k\sigma} C^\dagger_{\sigma} K^{\sigma}
\gamma_\sigma^{-2} (\cos\gamma_\sigma -1) - u^*_{k\sigma} \sigma
X^{2\bar\sigma } \gamma_\sigma^{-1} \sin\gamma,
\end{equation}
for transformed creation operators.

Then the transformed Hamiltonian for spin-down electrons has the
form
\begin{equation}\label{htildu}
\tilde H_\downarrow =  E_{d\downarrow} X^{\downarrow\downarrow} +
\sum_{kk'} \tilde \varepsilon^\sigma_{kk'}
a^\dag_{k\downarrow}a_{k'\downarrow}
\end{equation}
where $E_{d\downarrow}$ and $ \tilde \varepsilon^\sigma_{kk'}$ are
given by the same Eqs. (\ref{level}) and (\ref{tildec}) as in the
previous case, but with the energy $\varepsilon_d^{(2)}$
substituting for $\varepsilon_d$ and the correlation function
$K^\sigma$ taken from (\ref{com2}). It should be stressed that the
tunneling through the QD is impossible at zero source-drain bias
because of the spin blockade in spin-polarized electrodes. Only
spin-up carriers exist around Fermi level, and these electrons may
be injected into QD only when accompanied by the spin-flip
excitations given by the operators $X^{\downarrow\uparrow}$ in the
intermediate state with $N=1$ of cotunneling process. These
processes are inelastic and exponentially weak ($\sim V^2$ in
transparency and $\sim V^4$ in conductance). More detailed
discussion of spin-dependent tunneling is postponed till Section
\ref{TID}.

\subsection{Time-Dependent Transformation}
\label{TDP}

As was discussed above an experimental conditions may be created
when the tunneling matrix element $V_k(t)$ in the Hamiltonian
(\ref{ham1}) becomes time dependent. The canonical transformation
(\ref{Canon01}), as described in the previous section, cannot be
straightforwardly applied. Its generalization is in order.

We start with the temporal Schroedinger equation
\begin{equation}\label{Schro01}
{\cal L} \psi=0
\end{equation}
 and look for the time dependent transformation matrix ${\widetilde S}(t)$,
which transforms (\ref{Schro01}) into
\begin{equation}\label{Schro02}
{\widetilde {\cal L}}\widetilde \psi =0
\end{equation}
with transformed  operator
$$
{\widetilde {\cal L}}(t) =e^{S(t)}He^{-S(t)} -i\hbar
e^{S(t)}\frac{\partial}{\partial t}e^{-S(t)}
$$
for the new wave function $\widetilde \psi = e^{S(t)}\psi$.

The Hamiltonian $\widetilde H$ satisfying the equation
(\ref{Schro02}) can be written as
\begin{equation}\label{Schro04}
\widetilde{H} = e^SHe^{-S} + i\hbar\int\limits_{0}^1d\lambda
e^{\lambda S}\dot{S} e^{-\lambda S}.
\end{equation}
(see Appendix).

The Hamiltonian (\ref{Schro04}) contains now two terms of which
the first one is just a modification of the Hamiltonian
(\ref{Canon01}) of the time independent case. It means that all
the equations in Subsection \ref{TIT} hold except for Eq.
(\ref{coeff}), which defines the coefficients $u_{k\sigma}$ of the
canonical transformation. These coefficients must be now found
anew. There is also the second term in the right-hand side of Eq.
(\ref{Schro04}), which is responsible for non-adiabatic effects.

In order to find the canonical transformation parameters
$u_{k\sigma}$ we write explicitly the condition
\begin{widetext}
\begin{equation}\label{elimination_time}
u_{k\sigma} \gamma_\sigma^{-1} [(\varepsilon_d - Z_\sigma)
\sin\gamma_\sigma\cos\gamma_\sigma - ( \varepsilon_{k\sigma}
-Z_\sigma) \sin\gamma_\sigma + T_\sigma\cos^2\gamma_\sigma -
T_\sigma^*\sin^2\gamma - T_\sigma \cos\gamma_\sigma]
%$$$$
+ V_k \cos\gamma_\sigma =
$$$$
 - i\hbar\dot{u}_{k\sigma} \frac{\sin
\gamma_\sigma}{\gamma_\sigma} - i\hbar u_{k\sigma}
\sum\limits_{k'}\left[\dot{u}_{k'\sigma} u^*_{k'\sigma}
\frac{1}{2\gamma_\sigma^3} (\sin\gamma_\sigma\cos\gamma_\sigma +
\gamma_\sigma-2\sin\gamma_\sigma)\right.
%$$$$
+ \left.\dot{u}_{k'\sigma}^*u_{k'\sigma}
\frac{1}{2\gamma_\sigma^2}(1 -
\frac{\sin\gamma_\sigma\cos\gamma_\sigma}{\gamma_\sigma})\right] = 0
\end{equation}
\end{widetext}
of elimination of the QD - lead tunneling in the transformed
Hamiltonian. Here both the tunneling amplitude $V_{k\sigma}$ and
transformation parameter $u_{k\sigma}$ are functions of time. The
condition (\ref{elimination_time}) contains a number of terms with
the time derivatives $\dot{u}_{k\sigma}$. Neglecting these time
derivatives would correspond to the adiabatic approximation where
the variation of the tunneling amplitude is slow enough and the
whole electron system always have enough time to readjust to the
varying tunneling amplitude without additional level mixing. Then
one can check straightforwardly that Eq. (\ref{coeff}) with the time
dependent $V_k(t)$ solves Eq. (\ref{elimination_time}).

Now we carry out a more general analysis going {\it beyond} the
adiabatic approximation. For this sake we multiply Eq.
(\ref{elimination_time}) by $u^*_{k\sigma}$ and sum over $k$, which
leads to the equation
\begin{equation}\label{elimination_time_1}
(\varepsilon_d - Z_\sigma) \tan\gamma_\sigma  + T_\sigma -
T^*_\sigma\tan^2\gamma_\sigma =
$$$$
-i \hbar \frac{d}{dt} \tan\gamma_\sigma - i \hbar
\frac{\tan\gamma_\sigma}{2 \gamma_\sigma^2}
\sum\limits_{k'}(\dot{u}_{k'\sigma}
u^*_{k'\sigma}-\dot{u}_{k'\sigma}^* u_{k'\sigma}).
\end{equation}
Substituting  Eq. (\ref{elimination_time_1}) into Eq.
(\ref{elimination_time}) we get the equation
\begin{equation}\label{elimination_time_a}
u_{k\sigma}[(\varepsilon_{k\sigma} - Z_\sigma)\tan\gamma_\sigma +
T_\sigma] - \gamma V_{k\sigma} =
$$$$
-i\hbar \frac{\tan \gamma_\sigma}{\gamma_\sigma^2} u_{k\sigma}
\sum\limits_{k'} u_{k'\alpha'\sigma}^* \dot{u}_{k'\sigma} + i \hbar
\dot{u}_{k\sigma} \tan \gamma_\sigma .
\end{equation}
Neglecting the time derivatives of transformation parameter
$u_k(t)$ in the r.h.s of Eq. (\ref{elimination_time_a}), which
corresponds to the adiabatic approximation, yields Eq.
(\ref{coeff}) with the time dependent tunneling amplitude
$V_k(t)$. All the other equations obtained in Subsection \ref{TIT}
also hold. Accounting for the r.h.s. of Eq.
(\ref{elimination_time_a}) allows one to obtain nonadiabatic
corrections.

Having in mind that the quantities $\varepsilon_d$,
$\gamma_\sigma$ and $Z_\sigma$ are explicitly real, we separate
real and imaginary parts in Eq. (\ref{elimination_time_1}) and
thus get two equations
\begin{equation}\label{eltim2}
T_\sigma - T_\sigma^* = -2i\hbar \dot{\gamma_\sigma}
\end{equation}
and
\begin{equation}\label{eltim3}
(\varepsilon_d - Z_\sigma) \tan\gamma_\sigma  + \frac{T_\sigma +
T_\sigma^*}{2} (1 - \tan^2\gamma_\sigma) =
$$$$
- i \hbar \frac{\tan\gamma_\sigma}{2 \gamma_\sigma^2}
\sum\limits_{k'}(\dot{u}_{k'\sigma}
u^*_{k'\sigma}-\dot{u}_{k'\sigma}^* u_{k'\sigma}).
\end{equation}
These equations may be instrumental in looking for solutions
$u_{k\sigma}(t)$ for specific problems. It follows from
(\ref{eltim2}) that the quantity $T$ which was real for the time
independent case, remains real also in the adiabatic approximation
for the time dependent case.

Returning back to the transformed Hamiltonian (\ref{Schro04}), we
note that the first term $e^SHe^{-S}$ is now time-dependent due to
the time dependence of $S$. Carrying out the transformation in the
same fashion as in the previous section we get the time dependent
energy level
\begin{equation}\label{ed}
E_{d\sigma} = E_{d\sigma}^{(a)} + E_{d\sigma}^{(b)}
\end{equation}
where
\begin{equation}\label{eda}
E_{d\sigma}^{(a)} = \frac{\varepsilon_d + Z_\sigma
\tan^2\gamma_\sigma - (T_\sigma + T_\sigma^*) \tan\gamma_\sigma}{1
+ \tan^2\gamma_\sigma}
\end{equation}
is the same energy level (\ref{etild})  as before but with the
coefficients $Z_\sigma, T_\sigma, T_\sigma^*, \gamma_\sigma$
depending parametrically on time $t$. Thus, in adiabatic
approximation the time dependence of the resonance level position
$E_{d\sigma}^{(a)}(t)$  is determined by Eq. \ref{level}) with the
time-dependent self energy part
\begin{equation}\label{massopt}
L_\sigma(\varepsilon, t) = \sum_{k}
\frac{|V_{k}(t)|^2}{\varepsilon - \varepsilon_{k\sigma}}.
\end{equation}

There is also the nonadiabatic correction
\begin{equation}\label{edb}
E_{d\sigma}^{(b)}(t)= - \frac{1}{2} i \hbar
\frac{\sin^2\gamma}{\gamma^2}\sum\limits_{k\alpha}(
\dot{u}_{k\alpha\sigma}u^*_{k\alpha\sigma}
-\dot{u}_{k\alpha\sigma}^*u_{k\alpha\sigma} ).
\end{equation}
To calculate the time dependent coefficients, one has to specify the
form of tunneling amplitudes $V_k(t)$. An example of time-dependent
tunneling will be considered in the next section.

\section{Tunnel conductance}
\label{TC}

 In this section we study the tunnel conductance basing
on the transformed Hamiltonian $\widetilde H$. The tunnel current
may be calculated, e.g., by means of the Keldysh technique, where
the bias $eV$ is included in the zero order Hamiltonian and the
scattering $\sim W_{lk,rk'}$ is considered as a perturbation.
First, we calculate the spin-polarized current through an
immovable quantum dot and then discuss the modulation of this
current due to oscillatory motion of the dot.

\subsection{Tunneling through static dot}
\label{TID}

Far from the boundary between two adjacent charge sectors with
$N=1$ and $N=2$, where both the  levels
$\varepsilon_{d\uparrow}^{(1)}$ and
$\varepsilon_{d\downarrow}^{(2)}$ are far from the chemical
potential $\mu$, the Keldysh method applied to the Hamiltonian
(\ref{htildu}) in a single loop approximation gives the
conventional golden rule equation
\begin{widetext}
\begin{equation}\label{goldo}
I = e \frac{2\pi}{\hbar} \sum \limits_{kk'}\sum_\sigma
|W^\sigma_{lk,rk'}|^2\delta(\epsilon_{lk\sigma} + e\upsilon -
\epsilon_{rk'\sigma}) \{ f(\epsilon_{lk\sigma}) [1 -
f(\epsilon_{rk'\sigma})] - f(\epsilon_{rk'\sigma}) [1 -
f(\epsilon_{lk\sigma})] \}
\end{equation}
\end{widetext}
where $e\upsilon$ is the source-drain bias, $f(\epsilon_{k\sigma})$
is the Fermi distribution function for spin-polarized electrons, and
the scattering amplitude $W_{lk,rk'}$ is defined in Eq.
(\ref{scatamp-tit}).

The standard Coulomb diamond diagram for tunneling conductance
$G(v_g, e\upsilon)$ is distorted in the region of gate voltages
$v_g$ corresponding to the change of QD occupation $(N=1) \to
(N=2)$. In case of completely spin polarized dots the Coulomb step
corresponding to the resonance $E(N=1, v_g)=E(N=2, v_g)-\mu$ in
the current voltage characteristics is absent at zero bias  and
zero temperature due to the spin blockade. The chemical potential
$\mu$ is pinned to the Fermi level of spin up electrons
$\varepsilon_{F\uparrow}$, but the resonance level
$E_{d\downarrow}=E(N=2)-E(N=1)$ belongs to the down spin electron
(see Fig. \ref{f.spul}). Thus the tunneling at zero bias
$e\upsilon=0$ is suppressed by the spin blockade. This blockade
may be surmounted by means of finite source-drain bias
compensating the energy gap. However, the conditions for the onset
of spin up and spin down tunnel current are different.

\begin{figure}[h]
\begin{center}
  \includegraphics[width=4cm,angle=0]{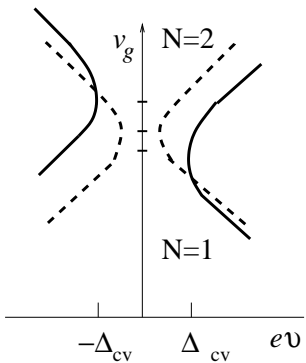}\hspace*{5mm}
\end{center}
\vspace*{-3mm}
  \caption{Coulomb diamonds diagram for tunneling conductance $G(v_g,e\upsilon)$
  near the border  $(N=1)/(N=2)$   for  down-spin and up-spin electrons
  (solid and dashed lines, respectively). Three bars on the $v_g$
  axis mark the values of the gate voltage corresponding to the
  level $E_{d\downarrow}$ in resonance with $\varepsilon_c$,
  $\varepsilon_F$, $\varepsilon_v$ (bottom-up)
 .
  }
  \label{f.diamond}
\end{figure}

The boundary of the Coulomb diamond with $N=1$ for spin down
electrons follows the evolution of the resonance level
$E_{d\downarrow}$ until this level approaches from above the
bottom of conduction band $\varepsilon_{c\downarrow}$. When it
crosses the band edge, the resonance tunneling is no more possible
at $eV < \Delta_{cv}$, so the Coulomb resonance line for $G(v_g,
e\upsilon)$ deviates from the linear behavior, as it is sketched
in Fig. \ref{f.diamond} (solid line). Negative-bias part of the
Coulomb diamond  corresponds to the hole tunneling through the
occupied resonance level in the sector $N=2$. It is distorted in
the same way in the region of $v_g$ where this level matches the
top of the valence band (second branch of the solid line). When
the down-spin level is deep enough in the conduction or valence
band, the linear behavior is restored again.

The blockade for spin up electrons near the boundary $(N=1)/(N=2)$
is lifted at $e\upsilon$ compensating the spin flip excitation in
the dot. As was mentioned above, the spin up electron tunneling is
allowed provided the dot is excited to the state $E_{d\downarrow}$
given by the solution of Eq. (\ref{level}) for the doubly occupied
dot, and the splitting energy may be estimated as
$\Delta_{\uparrow\downarrow}\approx
|L(E_{d\uparrow})-L(E_{d\downarrow})|$ (see Ref. \onlinecite{Pas}
for experimental determination of such splitting in the Kondo
tunneling regime). The line of resonance tunneling for spin-up
electrons is drawn in Fig. \ref{f.diamond} by the dashed curve.

Besides, the tunneling transparency is especially sensitive to the
position of the dot level in the near vicinity of band edges. To
investigate this dependence, let us find the explicit equation for
the tunnel conductance from Eqs. (\ref{goldo}) and
(\ref{scatamp-tit}). Changing summation over $kk'$ for integration
over $\epsilon\epsilon'$ in a usual way and performing standard
calculations, one gets the equation
\begin{equation}\label{conduct}
G_\downarrow(v_g,\Delta_{cv} ) =
$$$$
\frac{e^2}{h}\frac{\Gamma_l\Gamma_r}{(2\pi^2)}
    \frac{R^2(\gamma_\downarrow)}{|E_{d\downarrow}-\varepsilon_c|^2}
   \left[1+\frac{R(\gamma_\downarrow)L(E_{d\downarrow})}
   {\Delta_{cv}}
   \right]^2
\end{equation}
for the threshold value of $e\upsilon \to\Delta_{cv}+o$ in case of
2D electron gas in the planar leads. Here $\Gamma_i=2\pi V^2_iS_i$
is the tunneling rate for the lead $i$ obtained in the approximation
of $V_{ik}=V_i$ and constant electron density of states
$S_i(\varepsilon)=S_i$, which is valid for 2D electrons. It should
be taken into account that the definition of the position of the
resonance level $E_{d\downarrow}$ falling into the conduction band
continuum implies the procedure described below Eq. (\ref{level})
The function $L(\epsilon)$ has a logarithmic singularity at the band
edge, ${\rm Re}\,L(\epsilon \to \varepsilon_c) \sim
-\ln(|\epsilon-\varepsilon_c|/\Delta_{cv})$. As a result, Eq.
(\ref{level}) has either one or two solutions depending on a
position of the level $\varepsilon_{d\downarrow}^{(2)}-v_g$ relative
to the band edge.\cite{HA76}

The resonance factor $|E_{d\downarrow}-\varepsilon_c|^2$ in the
denominator and singular factor $L(E_{d\downarrow})$ in the
numerator of Eq. (\ref{conduct}) result in noticeable enhancement
of $G$ at the boundary of Coulomb diamond.

\begin{figure}[h]
\begin{center}
  \includegraphics[width=6 cm,angle=0]{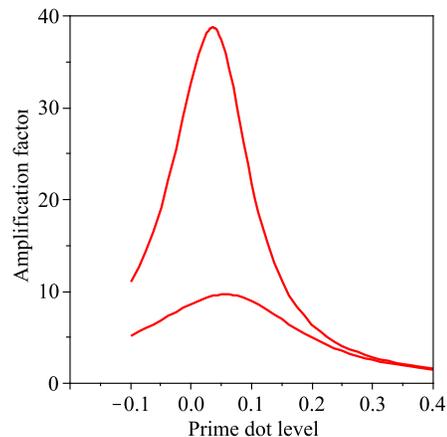}\hspace*{5mm}
\end{center}
\vspace*{-3mm}\caption{Amplification factor
$\rho(\varepsilon_d^{(2)})$ for two values of tunneling rate
$\Gamma/D=0.01$ (lower curve) and $\Gamma/D=0.02$ (upper curve).}
\label{f.amp}
\end{figure}

This enhancement is characterized by evolution of the ratio
\begin{equation}\label{enhan}
\rho(v_g) = \frac{R^2(\gamma_\downarrow)}{|E_{d\downarrow} -
\varepsilon_c|^2}
\end{equation}
from its value in the middle of Coulomb diamond to that in the
vicinity the point $\varepsilon_d(v_g) = \varepsilon_c$. Here and
below we omit the superscript (2) in the notation of
$\varepsilon_d$. Far from the Coulomb resonance the difference
between $|E_{d\downarrow}|$ and $\varepsilon_{d}$ is small and
$\tan^2\gamma_\downarrow \sim V^2/(\varepsilon_{d}-\varepsilon_c)
\ll 1$. The function $R(\gamma_\downarrow)$ tends to 1/2 in this
limit, so that the factor $\rho$ may be estimated as $\rho \approx
1/4(\varepsilon_{d}-\varepsilon_c)$. Near the band edge
$\varepsilon_c$ the factor $R(\gamma_\downarrow)\approx
\cot\gamma_\downarrow=\sqrt{|E_{d\downarrow}-\varepsilon_c|}/\Gamma_r$
[see Eqs. (\ref{deriv}),(\ref{factor})], so that  $\rho \sim
[|E_{d\downarrow}-\varepsilon_c|\Gamma_r]^{-1}$. Numerical estimates
of this enhancement are presented at Fig. \ref{f.amp}. In these
calculations the density of states was assumed to be constant in the
lower part of 2D conduction band, so that the self energy in the
r.h.s. of Eq. (\ref{level}) may be approximated as $L(E)\approx
\Gamma_r {\rm Ln}(E/D)$, where $D$ is the effective width of
conduction band, and the argument $E$ is complex. Then the solutions
$E=E_{d\downarrow}$ of Eq. (\ref{level}) are expressed via the
Lambert W-function $W(n,x)$
\begin{equation}\label{lambert}
E_{d\downarrow}= -\Gamma_r W(n,
-{\Gamma_r^{-1}}{e^{-{\varepsilon_d}/{\Gamma_r}}})
\end{equation}
(here the reference point is $\varepsilon_c=0$, all energy
parameters are measured in units $D$, index $n$ enumerates branches
of W-function). There are two solutions for $E_{d\downarrow}$ near
the band edge.\cite{HA76} The principal branch $n=0$ gives the
discrete level in the gap, and the branch $n=1$ corresponds to the
resonance in the band. Here we are interested in the latter state.

The amplification reaches its maximum when the difference
$|E_{d\downarrow}-\varepsilon_c|$ comes up with $\Gamma_r$,
therefore the smaller $\Gamma_r$, the bigger is the enhancement
factor $\rho$. The factor $RL/\Delta_{cv}$ in the square brackets
in Eq. (\ref{conduct}) behaves as $\epsilon^{1/2}\ln \epsilon$ in
the vicinity $\epsilon=|E_{d\downarrow}-\varepsilon_c|$ of the
band edge and thus gives additional contribution to this
enhancement. Both analytical and numerical estimates confirm this
statement. Experimentally, this effect should be observed as an
increase of tunneling conductance on the boundary of Coulomb
diamond in the vicinity of the threshold value of $e\upsilon_{\rm
th} =\Delta_{cv}$. A similar effect should arise on the hole side
of the Coulomb diamond diagram ($e\upsilon_{\rm th}
=-\Delta_{cv}$), where the occupied level $E_{d\downarrow}$
crosses the top of the down spin valence band (solid lines in Fig.
\ref{f.diamond}).

\subsection{Tunneling through moving dot}
\label{TMD}

As is shown above, the canonical transformation allows one to
distinguish between the adiabatic and non-adiabatic contributions to
the tunneling amplitude. Let us first discuss the adiabatic
corrections to the inelastic current given by Eq. (\ref{goldo}) and
illustrated by Fig. \ref{f.amp}. In the vicinity in the point
$(N=1)/(N=2)$ of the Coulomb diamond diagram the adiabatic position
of the down spin level $E_{d\downarrow}^{(a)}$  given by the
solution of equation
\begin{equation}\label{ead}
E_{d\downarrow}^{(a)}(t)= \sum_{j,k}
\frac{|V_{jk}(t)|^2}{E_{d\downarrow}^{(a)}(t) - \epsilon_{jk}}
\end{equation}
rocks around $\epsilon_{cr}$ or $\epsilon_{vl}$ (see Fig.
\ref{f.spul}). This solution depends parametrically on time via
the oscillating tunnel coupling $|V_{jk}(t)|^2$. The above
analysis of the static case prompts that time-dependent
corrections become significant at finite bias close to the
threshold value $e\upsilon_{\rm th} = \pm \Delta_{cv}$. The
adiabatic evolution of the level $E_{d\downarrow}^{(a)}$ in time
near $\varepsilon_c$ is given by the same Eq. (\ref{lambert}),
where the tunneling rate $\Gamma_r$ parametrically depends on $t$.
If the time-dependent perturbation is weak in comparison with the
static value of tunneling rate, the temporal component of
$E_{d\downarrow}^{(a)}$ may be found perturbatively. Representing
the tunneling rate as $\Gamma_r = \Gamma_{r0} + \delta\Gamma(t)$
and expanding Eq. (\ref{lambert}) around the time-independent
value marked by index '0', we get
\begin{equation}\label{expand}
E_{d\downarrow}^{(a)}(t)=E_{d\downarrow}^{(0)} -\frac{W_0}{1 + W_0}
\left( \frac{\varepsilon_{d}}{\Gamma_{r0}} +
W_0\right)\delta\Gamma(t)
\end{equation}
When deriving Eq. (\ref{expand}), the equality $W(x)=x[1+W(x)]dW/dx$
is used.

This time dependence turns into the corresponding adiabatic time
dependence of tunnel conductance, mainly via the enhancement
factor $\rho(E_{d\downarrow})$ (\ref{enhan}). One may expect that
the slow adiabatic variations of $G(t)$ will be especially
distinct at $v_g$ corresponding to steep slopes of $\rho(t)$ (Fig.
\ref{f.amp}) at $e\upsilon \sim \pm \Delta_{cv}$ near the boundary
$(N=1)/(N=2)$ of the Coulomb diamond diagram (Fig.
\ref{f.diamond}).

\subsection{Weak time dependent perturbation}
\label{WTDP}

Calculation accounting for nonadiabatic corrections to tunnel
conductance is generally an extremely complicated nonlinear
problem. To make it tractable, we assume that the time dependent
part of $V_{jk\sigma}$ is only a small periodic perturbation with
respect to the time independent part $V_{jk\sigma}^{(0)}$ and
consider only the linear response given by the first harmonics.
The nonlinear effects will be discussed separately. This approach
allows one to pick out first nonadiabatic corrections to tunneling
amplitude $W_{kk'}$, which turns out to be small everywhere except
in the vicinity of band edges.

Let us assume that the tunneling integral has the form
\begin{equation}\label{hybridization}
V_{k}(t) = V^{(0)}_{k} + V^{(1)}_{k} \cos\omega t
\end{equation}
(here and below indices $j\sigma$ are omitted for the sake of
brevity). Here $V^{(0)}_{k} \gg V^{(1)}_{k}$. The solution of Eq.
(\ref{elimination_time}) is looked for in the form
\begin{equation}\label{solution}
u_{k} = u^{(0)}_{k} + u^{(1)}_{k} \cos\omega t + i v^{(1)}_{k}
\sin\omega t
\end{equation}
where the time dependent corrections to $u^{(0)}_{k}$ are also
small. Then we vary all the coefficients in Eq. (\ref{eltim2}) with
respect to $u^{(0)}_{k}=g_k V^{(0)}_{k}$ [see Eq. (\ref{coeff})] and
$V^{(0)}_{k}$ and collect separately all the terms containing
$\cos\omega t$ and $\sin\omega t$ respectively. Substituting then
$u_{k}$ in (\ref{scatamp}), we reduce the scattering amplitude
$W_{kk'}$ to the following form
\begin{equation}\label{scatamp-tdt1}
W_{kk'} = \overline{W}_{kk'} + W^{(1)}_{kk'} \cos\omega t +
\hbar\omega W^{(2)}_{kk'} \sin\omega t
\end{equation}
where the coefficients of the cosine and sine terms can be
explicitly calculated. The cosine coefficient
\begin{equation}\label{cosine}
W^{(1)}_{kk'} =  \sum_{q} \left[\frac{\delta
\overline{W}_{kk'}}{\delta V_{q}} V^{(1)}_{q} + \frac{\delta
\overline{W}_{kk'}}{\delta V^*_{q}} V^{(1)*}_{q} \right]
\end{equation}
is obtained by varying equation (\ref{scatamp-tit}) over the
hybridization potential $V_{q}$, whereas the sine coefficient
\begin{equation}\label{sine}
W^{(2)}_{kk'} =
$$$$
i \sum_{q} g_{q}^2 \left[ \frac{\delta W_{kk'}}{\delta u_{q}}
V^{(1)}_{q} - \frac{\delta W_{kk'}}{\delta u^*_{q}} V^{(1)*}_{q}
\right] + W^{(2),nonad}_{kk'}
\end{equation}
is obtained by varying equation (\ref{scatamp}) over the
transformation parameter $u_{q}$. After the variation $u_{k}$ in
the form Eq. (\ref{coeff}) can be substituted. Then
\begin{widetext}
$$
W^{(2),nonad}_{kk'} = - i \left[ \frac{V^{(1)*}_{k'} V^{(0)}_{k} -
V^{(1)}_{k} V^{(0)*}_{k'}}{\Delta_{k\sigma}\Delta_{k'\sigma}}
\widetilde{R}(\gamma_\sigma) + \frac{V^{(0)}_{k}
V_{k'}^{(0)*}}{2\Delta_{k\sigma}\Delta_{k'\sigma}}
\widetilde{R}^2(\gamma_\sigma)  \sum\limits_{k''}
\frac{V^{(1)}_{k''} V^{(0)*}_{k''} - V_{k''}^{(0)*}
V^{(0)}_{k''}}{\Delta_{k''\sigma}\Delta_{k''\sigma}} \right]
$$
\end{widetext}
with
$$
\widetilde{R}_\sigma = \frac{R_\sigma}{\sqrt{1 +
\tan^2\gamma_\sigma}}.
$$
%deriv

The most divergent terms in (\ref{cosine}) and (\ref{sine}) appear
when we vary only explicitly written $V^{(0)}_{k}$ in Eq.
(\ref{scatamp-tit}) and $u_{k'}$ in Eq. (\ref{scatamp}). As a
result keeping the leading terms we have
\begin{equation}\label{scatamp-tit-a}
W^{(1)}_{kk'} =
$$$$
(V^{(0)}_{k}V^{(1)*}_{k'} + V^{(0)*}_{k'} V^{(1)}_{k} )K^\sigma
\left( \frac{1}{\Delta_{k\sigma}} + \frac{1}{\Delta_{k'\sigma}}
\right)R(\gamma_\sigma)
\end{equation}
and
\begin{equation}\label{scatamp-tit-a}
W^{(2)}_{kk'} = - i (V^{(0)}_{k}V^{(1)*}_{k'} - V^{(0)*}_{k'}
V^{(1)}_{k} ) \times
$$$$
K^\sigma \left( \frac{1}{\Delta^2_{k\sigma}} +
\frac{1}{\Delta^2_{k'\sigma}}\right)R(\gamma_\sigma)
\frac{\gamma_\sigma}{\tan\gamma_\sigma}.
\end{equation}

The corresponding corrections to the tunneling transparency near the
conductance band edge behave as
\begin{equation}\label{conduct-1}
\delta G^{(1)}_\downarrow(v_g,\Delta_{cv} ) \approx \frac{e^2}{h}
\frac{\Gamma_l\Gamma_r^{(1)}}{(2\pi^2)}
\frac{R^2(\gamma_\downarrow)}{|E_{d\downarrow} - \varepsilon_c|^{2}}
\cos\omega t
\end{equation}
and
\begin{equation}\label{conduct-2}
\delta G^{(2)}_\downarrow(v_g,\Delta_{cv} ) \approx
$$$$
(\hbar\omega)\frac{e^2}{h} \frac{\Gamma_l\Gamma_r^{(2)}}{(2\pi^2)}
\frac{R^2(\gamma_\downarrow)}{|E_{d\downarrow} -
\varepsilon_c|^{5/2}} \frac{\pi}{2\sqrt{\Gamma_r}} \sin\omega t
\end{equation}
where
$$
\Gamma_r^{(m)} = 2\pi S_r \left[i^{(m-1)} V_r^{(1)} V_r^{(0)*} +
(-i)^{(m-1)} V_r^{(0)} V_r^{(1)*}\right]
$$
is the adiabatic correction to the tunneling rate for $m=1$ and the
correction due to weak non-adiabatic effect for $m=2$. Equation
(\ref{conduct-2}) uses the fact that
$$
\gamma_\downarrow \to \frac{\pi}{2},\ \mbox{and}\ \
\tan\gamma_\downarrow = \sqrt{- L'} \approx
\sqrt{\frac{\Gamma_r}{|E_{d\downarrow} - \varepsilon_c|}}
$$
for $E_{d\downarrow} \to \varepsilon_c$ when $|E_{d\downarrow} -
\varepsilon_c| \approx \Gamma_r$. We use here the approximation
$V^{(m)}_{k'} \approx V^{(m)}_r$ similarly to the one used in Eq.
(\ref{conduct}).

The term with the sine in equation (\ref{scatamp-tdt1}) causes a
phase shift between the oscillations of the hybridization parameter
and the resulting current through the dot. If we neglect the $k$
dependence of the coefficients (\ref{cosine}) and (\ref{sine}) this
phase shift can be readily found
$$\varphi = - \arctan \left(
\frac{\delta G^{(2)}_\downarrow(v_g,\Delta_{cv} )}{\delta
G^{(1)}_\downarrow(v_g,\Delta_{cv} )} \right).
$$
It is expected to be generally rather small due to the factor
$g_{k}\hbar\omega$, which is usually very small unless the dot level
approaches the band edge $\varepsilon_c$. However close to the edge
this factor diverges and results in an increasing phase shift,
$$
\varphi \approx - \arctan \frac{\hbar\omega
\Gamma_r^{(2)}}{\Gamma_r^{(1)} \sqrt{\Gamma_r |E_{d\downarrow} -
\varepsilon_c|}} \approx - \arctan \frac{\hbar\omega
\Gamma_r^{(2)}}{\Gamma_r \Gamma_r^{(1)}}
$$
As result the phase shift $\varphi$ may become essential at
$E_{d\downarrow} \to \varepsilon_c$.

The non-adiabatic corrections to tunneling conductance acquire the
simple form of phase shift in oscillating cosine function only
until the parameter $f_k=g_k\hbar\omega \sim \hbar\omega/\Delta$
is small and the perturbative approach is valid. With increasing
$f$ one may expect appearance of higher harmonics $n\omega$ in
oscillating conductance. With further increase of the amplitude
$V^{(1)}$ the language of quasi-energy levels\cite{Z73} is more
appropriate. We plan to discuss it elsewhere.

\section{Concluding remarks}

We have discussed in this paper a new approach to the Anderson model
for a half-metallic electron liquid in a tunnel contact with a
moving nano-shuttle under strong Coulomb blockade. It is shown that
in the situation where the spin-flip cotunneling processes are
suppressed at low energies, the exact canonical transformation
eliminating the tunneling term in the Anderson Hamiltonian exists
even in the presence of strong Hubbard repulsion in the shuttle and
time-dependent lead-shuttle tunneling. This canonical transformation
in principle allows one to sort out the slow adiabatic
renormalization of the energy levels and the tunnel transparency and
to consider non-adiabatic corrections at least perturbatively. One
may also include the inelastic spin-flip processes in the canonical
transformation in the 4-th order of perturbation theory in $V_i$,
but these weak corrections do not change the above qualitative
picture. We also have calculated the weak non-adiabatic corrections
to tunneling transparency in a specific model of periodic sinusoidal
motion of the shuttle. Undoubtedly, strong non-adiabatic effects
should be taken into account in a more refined scheme: in the case
of a periodic time-dependent perturbation one should appeal to the
Floquet theorem in the time domain and use the quasi energy
language.\cite{Z73,Baron77}

Because of the energy gap for spin-flip processes in the
half-metallic leads the Kondo cotunneling processes are suppressed,
and the zero-bias anomaly in the tunnel conductance is absent.
Physical manifestations of the shuttling mechanism under discussion
arise in the form of time-dependent enhancement of conductance at
finite bias near the boundaries of Coulomb diamonds on the phase
diagram $G(v_g, e\upsilon)$. The boundaries themselves are distorted
due to the complete spin polarization of carriers (see Fig.
\ref{f.diamond}), and even the Coulomb blockade step corresponding
to occupation change from odd to even number of electrons in the dot
is absent at zero bias. One may expect appearance of quasi energy
satellites in this part of the phase diagram $G(v_g, e\upsilon)$
when the shuttle motion is essentially non-adiabatic. This regime is
a subject for future studies.

\appendix

\section{Time dependent canonical transformation}

To diagonalize the Schr\"odinger equation (\ref{Schro02}) for the
wave function ${\widetilde \psi}=e^{S(t)}\psi$, we conjecture the
form
$$\widetilde{H} = e^SHe^{-S} + S_1$$ for the canonically transformed
Hamiltonian. Then substituting $\widetilde{H}$ and
$\widetilde{\psi}$ into Eq. (\ref{Schro02}), we obtain
\begin{equation}\label{Schro03}
i\hbar\left(\frac{\partial}{\partial t}e^S\right)\psi + i\hbar e^S
\frac{\partial\psi}{\partial t} = (e^SHe^{-S} + S_1)e^S\psi~.
\end{equation}
The time derivative of exponent is found by means of the operator
equation
\begin{equation}\label{Schro05}
\frac{d}{dt}e^S = \int \limits_0^1 e^{\lambda S}\dot{S}
e^{-\lambda (S-1)} d\lambda   = e^S \left[ \int\limits_0^1
e^{-\lambda S}\dot{S} e^{\lambda S}d \lambda\right].
\end{equation}
Then using (\ref{Schro05}) and (\ref{Schro01}) in (\ref{Schro03})
yields
$$
i\hbar \left[\int\limits_{\lambda=0}^1 d\lambda e^{\lambda
S}\dot{S}e^{-\lambda S}\right] e^S\psi + e^S
H\psi=e^SH\psi+S_1e^S\psi.
$$
from where we straightforwardly get $S_1$ and Eq. (\ref{Schro04}).
\section{Tunneling amplitude for weak periodic potential}

To derive the first non-vanishing time-dependent corrections to
tunneling procedure, we vary the terms in Eq. (\ref{eltim2}). This
variation procedure results in the two linear equations
\begin{widetext}
\begin{equation}\label{seteq}
\begin{array}{l}
{u^{(1)}_k}^* V^{(0)}_{k} - u^{(1)}_{k} {V^{(0)}_{k}}^* -
\hbar\omega g_k \left({V^{(0)}_k}^* v^{(1)}_{k} - {v^{(1)}_k}^*
{V^{(0)}_k}\right) = - g_k \left({V^{(0)}_k}^* V^{(1)}_{k} -
V^{(0)}_k {V^{(1)}_{k}}^*\right),
\\
 \\
{v^{(1)}_k}^* V^{(0)}_k + {V^{(0)}_k}^* v^{(1)}_k + \hbar\omega
g_k ({V^{(0)}_k}^* u^{(1)}_k + {u^{(1)}_k}^* V^{(0)}_k) = 0
\end{array}
\end{equation}
\end{widetext}
where $g_k$ is explicitly real as defined in Eq. (\ref{coeff}).

The first of these equations contains only explicitly imaginary
terms whereas the second one contains only real terms. Therefore
we have only two equation for four real variables (two complex
variables). It is readily verified that
\begin{equation}\label{solution_p}
\begin{array}{c}
\overline{u}_k^{(1)} = \displaystyle\frac{g_k}{1 - (\hbar\omega
g_k)^2} V^{(1)}_k
\\
\overline{v}_k^{(1)} = - \displaystyle\frac{\hbar\omega g^2_k}{1 -
(\hbar\omega g_k)^2} V^{(1)}_k
\end{array}
\end{equation}
is a particular solution of the set of equations (\ref{seteq}).
One can also see that in the limit $g_k\hbar\omega \to 0$ this
solution reduces to the adiabatic approximation where we can take
the results obtained in Subsection \ref{TIT} for the time
independent problem and substitute there the tunneling amplitude
(\ref{hybridization}) slowly varying in time.

The general solution of the corresponding homogenous set of
equations reads
\begin{widetext}
\begin{equation}\label{solution_h}
\begin{array}{c}
\underline{u}^{(1)}_{k}  = - \frac{1}{2} \left(\hbar\omega g_k +
\displaystyle \frac{1}{\hbar\omega g_k}\right) (C_{1k} + iC_{2k}) +
\left(\hbar\omega g_k - \displaystyle \frac{1}{\hbar\omega
g_k}\right) \displaystyle \frac{V^{(0)}_{k}}{2
{V^{(0)}_{k}}^*}(C_{1k} - iC_{2k}),
\\
\\
\underline{v}^{(1)}_k = C_{1k} + iC_{2k}
\end{array}
\end{equation}
\end{widetext}
It should be added to the particular solution (\ref{solution_p}).
The parameters $C_{1k}$ and $C_{2k}$ are meanwhile arbitrary. These
parameters determined from the original cancelation equations
(\ref{elimination_time}) - (\ref{eltim3}) turn out to me small,
$C_1,\ C_2 \sim (g_k\hbar\omega)^3$ and  may result at most in
nonadiabatic corrections $\sim (g_k\hbar\omega)^2$.

In order to calculate the tunnel current we need the scattering
matrix element
\begin{equation}\label{scatamp-tdt}
W_{kk'} = W^{ad}_{kk'} + W^{nonad}_{kk'}
\end{equation}
where the first term $W^{ad}_{kk'}$ is obtained by substituting
the solution (\ref{solution}), (\ref{solution_p}) into Eq.
(\ref{scatamp}) and keeping the terms linear in $g_k\hbar\omega$.
This contribution remains finite in the adiabatic limit
$g_k\hbar\omega \to 0$. However it contains also nonadiabatic
corrections due to the third term in (\ref{solution}). These
corrections are determined by the equation
\begin{eqnarray}\label{scatamp-nad}
&& W^{nonad}_{kk'} = i\hbar (\dot{u}_{k'}^*u_{k} -
\dot{u}_{k}u^*_{k'}) \frac{1-\cos\gamma}{\gamma^2} \\
&& + i\hbar u_{k} u_{k'}^* \frac{(\cos\gamma-1)^2}{2\gamma^4}
\sum\limits_{k''} (\dot{u}_{k''}u^*_{k''} -
\dot{u}_{k''}^*u_{k''}) . \nonumber
\end{eqnarray}
which contains derivatives $\dot u_k$ and, hence, is purely
nonadiabatic, i.e. disappears in the limit $g_k\hbar\omega \to 0$.
Keeping only  terms linear in $g_k\hbar\omega$ in the scattering
matrix element (\ref{scatamp-tdt}), we come to Eq.
(\ref{scatamp-tdt1}).

\end{document}